\documentclass[twocolumn,showpacs,floatfix,prb,10,aps]{revtex4-1}
\usepackage{amsmath,amssymb}
\usepackage{epsfig,psfrag}
\usepackage{dcolumn}
\usepackage{bm}
\usepackage{graphicx}
\usepackage{color}

\def\wx{\omega_{\rm ex}}
\def\with{{\rm with}}

\def\and{{\rm and}}
\def\hn{\hat{n}} 
\def\bb{\hat{b}}
\def\bbd{\hat{b}^\dagger}
\def\d{\hat{d}}
\def\dd{\hat{d}^\dagger}
\def\cc{\hat{c}}
\def\cd{\hat{c}^\dagger}
\def\kb{k_{\rm B}}

\def\L{{\cal L}}

\def\T{{\cal T}}
\def\Tr{{\rm Tr}}

\def\b{{b}}

\def\hbar{\hslash}

\begin{document}

\title{Antiresonant quantum transport in ac driven molecular nanojunctions}
\author{V. Leyton$^{1}$, S. Weiss$^{2}$, and M. Thorwart$^{3}$}
\address{ $^1$ Facultad de Ciencias B\'asicas, Universidad Santiago de
	Cali, Calle 5 No. 62 - 00 Cali, Colombia
	\\ $^2$ Theoretische Physik, Universit\"at Duisburg-Essen \& CENIDE,
	D-47048 Duisburg, Germany
	\\ $^3$ I. Institut f\"ur Theoretische Physik, Universit\"at
	Hamburg, Jungiusstraße 9, D-20355 Hamburg, Germany } 
\date{\today}

\begin{abstract}
	We calculate the electric charge current flowing through a vibrating 
        molecular nanojunction, which is driven by an ac voltage, in its regime of nonlinear oscillations. 
        Without 
        loss of generality, we model the junction by a vibrating molecule which is doubly clamped to two metallic leads which are 
        biased by time-periodic ac voltages.  Dressed-electron
        tunneling between the leads and the molecule drives the
        mechanical degree of freedom out of equilibrium. In the deep
        quantum regime, where only a few vibrational quanta are excited, the formation of coherent
        vibrational resonances affects the dressed-electron tunneling. In turn, back action modifies the 
        electronic ac current passing through the junction. The concert of nonlinear vibrations and ac driving induces quantum transport currents 
        which are antiresonant to the applied ac voltage. 
        Quantum back action on the flowing nonequilibriun current allows us to obtain rather sharp spectroscopic information on the population of the mechanical vibrational states. 
\end{abstract}
%
%
\maketitle

\section{Introduction}

Fascinating progress has been achieved in downsizing artifically made
condensed-matter devices. The study of micromechanical systems has evolved towards 
nanoelectromechanical systems (NEMS) that are at the core of molecular scale
electronics\cite{Ferdin2017}. Thereby, the fundamental physical
limits set by the laws of quantum mechanics are rapidly
approached. The ultimate potential for nanoelectromechanical devices
is governed by the ability to detect motional response to various
external stimuli giving a variety of physical phenomena including
electronic correlations \cite{Choi2017} as well as magnetism and other
spin-related effects \cite{Kiran2017,Gaudenzi2017,Gaudenzi2017}.  Molecular
vibrations \cite{Kocic2017} and junction mechanics \cite{Hybertsen2017}
are also under consideration in view of their thermal properties \cite{Cui2017,Tan2017}. 

Several experimental realizations of nanoscale systems  exist which display mechanical 
vibrations, such as, for instance, transversely vibrating nanobeams or
lithographically patterned doubly clamped suspended beams
\cite{LaHaye04}. Also suspended doubly clamped
carbon nanotubes exhibit a rich mechanical vibrational spectrum \cite{Safavi2012}.
Applications as electrometers\cite{Cleland1996,Cleland1998} for
detecting ultrasmall forces and displacements \cite{LaHaye04,Beil00}
have been reported.  Also NEMS are used for radio-frequency signal
processing \cite{Nguyen99} and chemical sensoring
\cite{Mohanty08,Mohanty06}. Other NEMS application include 
signal amplification in ultrasmall devices \cite{Mohanty07a,Mohanty07b} and spin readout
techniques \cite{Mohanty04}.  Fundamental physical phenomena emerge in
NEMS due to the interplay of electronic and mechanical degrees of
freedom, often immersed into a nonequilibrium environment 
\cite{Shekhter06,Regal08,Hansen2017}.

Due to their size, NEMS are of interest when studying the crossover from
the classical to the quantum regime, where quantum fluctuations in
transverse vibrations may drastically influence the 
dynamics \cite{Carr01,Carr01a,Werner04}. The possibility of observing macroscopic
quantum coherence is viable, since the quantized mechanical motion (phonons) involves a macroscopic number
of particles forming the nanobeam. Yet, coherence is significantly disturbed by the
interaction with the environment resulting in damping and
decoherence \cite{Imboden14}. Experiments have reported 
measurements of the nonlinear response of a radiofrequency mechanical
resonator which allows to obtain precise values of relevant mechanical
parameters of the resonator \cite{Aldridge05}, as well as the cooling of the
resonator motion by parametric coupling to a driven microwave-frequency 
superconducting resonator \cite{Rocheleau2010}.  

Most techniques used to detect and actuate NEMS in view of the quantum behavior
of their motion address linear response properties of transverse vibrations around
their eigenfrequencies. 
In order to measure the response to various external stimuli, experiments 
require an increased resolution of the position measurement
to the sub-thermal state
\cite{Beil00,LaHaye04,Poggio2007,Safavi2012,MacQuarrie2017}. As the
response of a damped linear quantum oscillator has a
Lorentzian shape, similar to a damped linear classical oscillator
\cite{Weiss1993,Chan2011}, a unique identification of the quantum
behavior of a nanoresonator in the regime of linear vibrations 
is sometimes hard to perform.

Interestingly, pronounced quantum features arise when driven damped nonlinear quantum 
resonators are considered. These are typically induced by the interplay 
of the nonlinearity and the external periodic driving \cite{peanoPRB,peanoCP,peanoNJP,peanoJCM1,peanoJCM2,Vicente1,Vicente2}.  
In the case of a driven dissipative quantum oscillator with a 
quartic nonlinearity, the oscillation amplitude in the steady state shows distinct 
quantum antiresonances for particular values of the driving frequencies 
\cite{peanoPRB,peanoCP,peanoNJP}. At those values, 
multiphoton transitions occur which are accompanied by a phase slip of the response relative 
to the excitation, such that an antiresonant line shape of the response and a 
driving induced dynamical bistability arise. Similar response characteristics is generated in a 
quantum mechanical two-level system which is coupled to a harmonic oscillator in the presence of 
driving (either of the two-level system or the oscillator) \cite{peanoJCM1,peanoJCM2}. 
This driven dissipative Jaynes-Cummings model is also intrinsically nonlinear, leading to a comparable  
response in terms of quantum antiresonances. Since these antiresonances are associated 
to multiphoton transitions, they are in general very sharp. Hence, it has been proposed to 
use them for the state detection of quantum bits \cite{Vicente1}. In fact, the sharpness of the 
antiresonances also leads to interesting quantum noise properties of the multiphoton 
transitions \cite{Vicente2}. Yet, the detection of the sharp antiresonances remains difficult experimentally. 
This motivates the study of these effects in quantum transport setups, as suggested in the present work.
 
Further reduction in size from NEMS to molecular electronics has been 
pursued during recent years. From the experimental point of view,   
transport setups have the advantage that
the current-voltage characteristics is accessible. Thus, it is
an interesting question to search for nonlinear features in the mechanical 
motion of vibrating molecules. Important vibrational effects in the quantum transport in  
  molecules concern phonon-assisted transport or non-linear vibrations \cite{Nitzan01,Nitzan,molel,Nitzan07,May04}, for a comprehensive
review of vibrational effects in molecular transport, see
Ref.\ \onlinecite{Nitzan07}. Cizek, Thoss, and Domcke \cite{Thoss04} treat
the inelastic regime by an electron-molecule scattering theory.
 In Ref.\ \onlinecite{Flensberg03}, one
vibrational mode has been investigated under the assumption of a
strong electron-phonon coupling, which gives rise to rather strong tunneling
broadening of the vibrational sidebands. A subsequent
work \cite{Braig03} included additional damping of the vibrational mode.
Vibrational effects in molecular transistors in the regime of 
sequential electron tunneling have also been
investigated in Ref.\ \onlinecite{Weiss2015}. 
Recently, implicit driving of the mechanical degrees of freedom induced by the
electronic current has been revealed \cite{Jin2015}. 
The electrons which tunnel through a voltage-biased tunnel junction drive a transmission line resonator 
out of equilibrium. Further, an external periodic bias voltage can modify the distribution of molecular
vibrations and the fluctuations of the molecular displacement \cite{Ueda2017}. Moreover, the
emission noise of a conductor can drive the state of a single-mode cavity
coupled to a voltage-biased quantum point contact \cite{Mendes2016}. 
When the molecular bridge has a permanent magnetic moment and a sizable magneto-mechanical coupling, the concept of nanocooling has been developed recently 
 \cite{Brueggemann2014,Brueggemann2016}, in which a spin-polarized electronic current is used to locally control the magnetic moment which may reduce the thermal population of the mechanical vibrational mode and thus cool it.

Current-induced non-equilibrium vibrations in single molecule devices
have been investigated in Refs.\ \onlinecite{Koch05,Ueda2016,Ueda2017},
again in the incoherent regime.  The impact of external light fields on
electronic transport has been analysed in detail in Ref.\ \onlinecite{Kohler05}. 
Moreover, charge transport through a vibrating molecule has been studied in terms of  Keldysh
Green's function perturbatively in the electron-phonon
coupling \cite{Ueda2017,Egger07}. Also nonequilibrium phonon dynamics in nanobeams and the related
phonon-assisted losses have been investigated in
Ref.\ \onlinecite{Haertle2015}.  

Antiresonances in quantum transport set-ups do not only occur when mechanical vibrations are present. In general, they can arise whenever nonlinear elements in a transport geometry occur. For instance,  antiresonances in the conductance of a ferromagnetic lead with a side-coupled quantum dot can occur on the level of a treatment in terms of the Landauer formula due to interference of a resonant and a nonresonant transport path through the system \cite{FengPhysica2005}. Likewise, when several quantum dots are arranged in different geometries (in series, in parallel, etc.), the intrinsic transport features also become nonlinear and resonances and antiresonances arise the current-voltage spectrum \cite{JAP2009}.  
Also, Fano-type antiresonances occur in the linear conductance as a function of the gate voltage in a multi-dot set-up when the tunneling coupling between the dot system and the leads is asymetric \cite{JAP2012}. Yet, the occurrence of antiresonances in a mechanically vibrating nanojunction has not been discussed so far in the literature.

\begin{figure}[t]
	\centering
        \includegraphics*[width=0.4\textwidth]{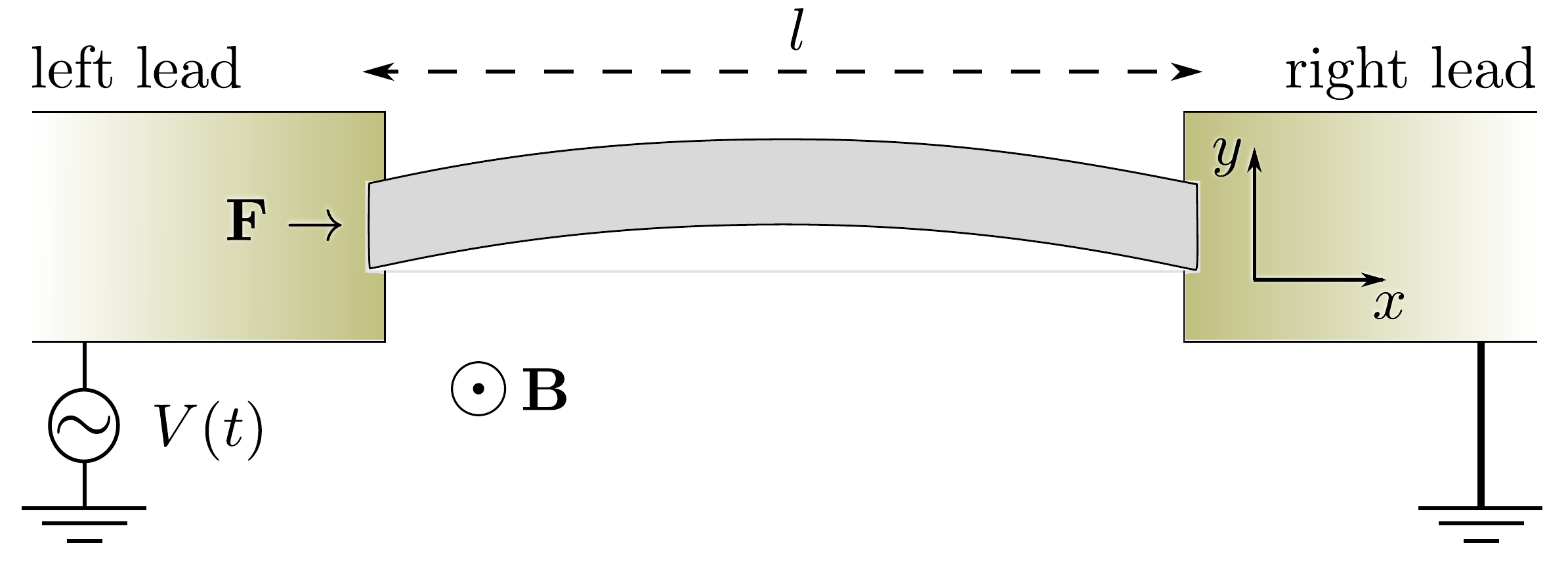}
	\caption{Sketch of a suspended nanobeam of length $l$
          clamped between two metallic leads.  We consider a constant external
          force $ {\bf F}$, applied in the longitudinal direction of
          the nanobeam (along the $x$-axis). Additionally, in order to
          control the electromechanical coupling, we assume a transversal 
          external magnetic field ${\bf B} = B_0 \hat{e}_z$.  The system
          is driven out of equilibrium by a time-dependent bias
          voltage $V(t)$.  \label{fig1}}
\end{figure}

In this work, we are interested in the interplay of nonlinear molecular vibrations and 
an external ac driving, in particular in the deep
quantum regime. We shall consider a molecular junction where its mechanical degree of freedom is described
by a monostable nonlinear oscillator with a Kerr nonlinearity, while its electronic degree of freedom is 
modelled as a single electronic level (quantum dot approximation). This carries the electrons tunneling through the system to two noninteracting electronic leads. In addition, a
periodically modulated bias voltage in the leads is considered in
order to drive the system out of equilibrium, [cf. Fig.~\ref{fig1}]. We consider the
regime of weak electromechanical interaction in which independent single-electron
tunneling processes between the leads and the junction modulate the junction's 
mechanical motion and may induce few-phonon transitions. We  
in particular identify the signatures of the nonlinear vibrations in the charge 
current flowing through the nanobeam. Moreover, we find 
 Fano-shaped resonances in the current, which are, in fact, antiresonances and which can be traced back to nonlinear resonances in the mechanical quantum dynamics. 

In Sec.~\ref{sec:model}, we introduce the Hamiltonian model.  The rotating wave approximation is
invoked and a time-dependent effective Hamiltonian is derived in Sec.\ \ref{sec:effHam}.  
Within a quantum master equation approach, discussed in Sec.~\ref{sec:masterequation}, we evaluate the current 
in the rotating as well as in the laboratory frame. We present our results in Sec.~\ref{sec:quantumosci}.  Finally we
conclude our findings in Sec.~\ref{sec:conclusions}.

%
%
\section{Model of an ac-driven nonlinear nanojunction}
\label{sec:model}
The setup of the molecular nanojunction depicted in Fig.~\ref{fig1} includes a
suspended nanobeam of length $l$, doubly clamped to normal conducting leads in the presence of a time-dependent
electrostatic potential $V(t)$.  An external magnetic field ${\bf B}$ is applied
perpendicular to the longitudinal axis of nanobeam to couple the mechanical motion to the
electronic degrees of freedom. Electrons can tunnel from the leads into a single electronic level of the nanobeam, which is assumed to form a quantum dot. The Hamiltonian is
\begin{equation}\label{eq:genmod}
H(t) = H_e + H_{em} + H_{m} + H_t +H_{leads}(t) \, .
\end{equation}
Here, $H_{e}$ is the Hamiltonian for the electrons passing through
the nanobeam, $H_{em}$ describes the coupling between the electronic and the 
mechanical degree of freedom, and 
$H_{m}$ contains the mechanical degree of freedom with 
nonlinear bending deflections induced by an external force
${\bf F} = F \hat{e}_x$ in longitudinal direction. The tunneling of electrons from (to) the leads
is accounted for by $H_{t}$ and $H_{leads}(t)$ covers the dynamics of
noninteracting electrons in the leads.

For the electron dynamics, we consider only one longitudinal 
energy state with energy $E$, associated with the motion of electrons along the
nanobeam, yielding
\begin{equation}   
H_e = E \, \dd \d \ ,
\end{equation}
where $d^\dagger$ ($d$) creates (destroys) an electron on the nanobeam.

For the mechanical dynamics, we want to consider the effect of a nonlinear vibrational mode of the nanojunction. The nonlinearity stems from the double clamping to the mechanical oscillator. To illustrate this in principle, we may consider a doubly clamped mechanical nanobeam. A nonlinear term can be easily obtained \cite{Werner04} by a constant longitudinal 
external force ${\bf F} = F \hat{e}_x$ with $F$ being close to the Euler buckling 
instability $F_c = {\cal E I }(\pi/l)^2$, with $\cal E$ being Young's
modulus of elasticity and $\cal I$ the area momentum of inertia. Close to this unstable point, 
the fundamental mode vanishes and higher modes and nonlinear effects become relevant. Then, the bending
deflections can be modeled by a single nonlinear vibrational mode
\cite{Werner04} (we set $\hbar = 1$)
\begin{equation}\label{hamm}
H_m = \omega_0 \, \hn +  \nu \, (\bbd +   \bb)^4/12 \  ,
\end{equation}
where
\begin{eqnarray}
\omega_0 &=& 4 \left( \frac{\pi}{l} \right)^2 \left[ \frac{\cal
    EI}{3\rho_b} \cdot \frac{F_c - F}{F_c} \right]^{1/2} \ , \\[0.4cm]
\nu &=& \frac{4F_c - F}{{\cal EI}\rho_b \, l} \cdot \frac{F_c}{F_c -
  F} \ ,
\end{eqnarray}
are the fundamental frequency of the bending mode and the Kerr
nonlinearity, respectively. Above, we have denoted by $\hn = \bbd \,
\b $ the phonon number operator.  Intrinsically, the bending
deflections affect the electronic dynamics through a very weak
electromechanical coupling that depends on an even power of the
nanobeam's deflection amplitude \cite{Weick2010,Rastelli2012}.   
This coupling is enhanced by the application of an external magnetic field.
Hence, the electromechanical coupling is tunable, where the electromagnetic
force exerting on the electrons depends on the bending of the nanobeam. 
For the sake of simplicity, we consider the magnetic field applied in the $z$-direction 
perpendicular to the nanobeam's longitudinal axis [cf.\ Fig.\ \ref{fig1}]. It has been shown in Ref.~\onlinecite{Rastelli2012} 
that the resulting electromechanical coupling might be written as 
\begin{equation} \label{eq:elemech}
  H_{em} = i \omega_0 \phi (\bbd  - \bb) \dd \d \, , 
\end{equation}
where the dimensionless coupling constant is given by
\begin{equation}\label{eq:elecoupling}
  \phi = \frac{\pi B_0 Y_0 l}{ \Phi_0} \int_0^l ds \, \frac{u_0(s)}{l} \, .
\end{equation}
Here, the magnitude of the external magnetic field is denoted by $B_0$, $Y_0$ is the amplitude of the zero point motion of 
the oscillator, the magnetic flux $\Phi_0=2\pi/e$ and $u_0 (s)$ is the profile of the
fundamental bending mode normalized\cite{Rastelli2012} according to $\int_0^l ds u_0^2(s)/l=1$. 

Instead of considering a nanobeam, a linear molecule, e.g., a carbon nanotube, can be used in a doubly clamped configuration and excited to its nonlinear regime. For specific molecules, the mechanical modes can be determined numerically, but eventually lead to a model in the form discussed above.

The tunneling coupling between the system's electronic state and the conducting leads is
provided by the tunneling Hamiltonian
\begin{eqnarray}\label{eq:tunneling}
  && H_{ t} = \sum_{p\,=\, l, \, r } H_p^+ + H_p^- \ , \\ && {\rm with}
  \quad H^-_p = \sum_{k} T_{{ p},k}\, \d \ \cd_{p,k} ,\ \and \quad
  H^+_p = (H_p^-)^\dagger . \nonumber
\end{eqnarray}
Here, $\cc^\dag_{ p,k}, (\cc_{ p,k})$ creates (annihilates) an electron 
in the lead $p=l, r$.  The coupling strengths are characterized by 
$T_{p,k}$, which induce a finite lifetime $\tau$ for
electrons in the nanobeam. Hence, a broadening of width $\Gamma = 1/\tau$
is generated for the electronic level of the nanobeam.  In the standard wide-band limit
approximation, one can neglect the energy dependence of the tunneling constants, i.e., 
 $ T_{p,k} \rightarrow T_p$, and assumes a constant
level broadening $\Gamma \propto \sum_p|T_{p}|^2$. Later on, we study
the weak coupling limit where $ \Gamma \ll \Delta
E, \omega_0$, with $\Delta E$ being the spacing of the quantized energies in the beam.

The leads are described by noninteracting electrons in the presence of an ac voltage
$V(t) = V_0 \cos (\wx t)$. Here, $V_0$ is the magnitude of the ac-driving voltage and $\wx$ 
the corresponding driving frequency. The resulting
electrostatic potential difference renders the single-particle electronic energies
in each lead time-dependent, according to $E_{ p,k}(t) = E_{p,k} + e
V_{p}(t)$, with $V_{p=r,l}(t) = \pm V(t) /2$ and $E_{p,k}$ being the electronic 
energies in each lead $p$. This results in the Hamiltonian
\begin{equation}\label{eq:leads}
  H_{leads} (t) = \sum_{p,k} E_{p,k}(t) \cd_{p,k} \cc^{}_{p,k}.
\end{equation}

\subsection{Time-dependent transformation of the Hamiltonian}
It is convenient to transform the time dependence in
Eq.~\eqref{eq:leads} together with the coupling term
Eq.~\eqref{eq:elemech} to the tunneling term Eq.~\eqref{eq:tunneling} by a unitary transformation \cite{Shekhter06,Rastelli2012,Segal17}
\begin{equation}\label{eq:Ftrans}
  {\cal U}(t) = \exp \left[\sum_{k,p} \varphi_p(t) \cd_{p,k} \cc_{p,k} +
    i \phi (\bbd + \bb)\dd \d \right]\, .
\end{equation}
Here, $\varphi_{p}(t) = {\rm e} \int_0^t ds\, V_{p}(s) = ( v_p
/\wx)\sin(\wx t)$ is the phase accumulated by the bias voltage with $
v_{p} = \pm e V_0$.  The result of the transformed tunneling term
Eq.~\eqref{eq:tunneling} reads
\begin{equation}\label{eq:lastunn}
  H_t'(t) = \sum_{p} \exp[-ie\varphi_p(t) + i \phi (\bbd + \bb)] \,
  H^-_p + {\rm h.c.} \ ,
\end{equation}  
and, for Eq.~\eqref{eq:leads} we find
\begin{equation}
  H_{leads}' = \sum_{p, k} E_{p,k} \cd_{p,k} \cc_{p,k} \, . 
\end{equation}
Note that after the transformation, all time-dependent interactions, which influence
the resonator externally, are shifted to the time-dependent tunneling term
Eq.~\eqref{eq:lastunn}. The Hamiltonian can thus be rewritten as
\begin{equation}
  H(t) = H_e + H_m + H_t'(t) + H_{leads}' . 
\end{equation}

It is convenient to rewrite Eq. \eqref{eq:lastunn} as an expansion in
orthogonal polynomials
\begin{widetext}
\begin{eqnarray}\label{eq:tunnfull}
H_t'(t) &=& \sum_{p} \sum_{n m m'} \left[ J_n(v_p/\wx) e^{i n \wx t}
  \right] \left[ e^{-\phi^2/2} \frac{(i\phi)^{m+m'}}{m! m'!}
  (\b^{\dagger})^m \b^{m'} \right] H_p^{-} + {\rm h.c.} \, , \nonumber \\
\end{eqnarray} 
\end{widetext}

where we have used the Jacobi-Anger identity $\exp[i(v_p/\wx) \sin (\wx t)] = \sum_n
J_n(v_p/\wx) e^{i n \wx t}$ for the accumulated phase $\varphi_p(t)$ with
$J_n(z)$ being the $n$th ordinary Bessel function of the first kind. In addition, we
have used the identity $e^{i\phi(\bbd + \bb)} = e^{-\phi^2/2} e^{i
  \phi \bbd} e^{i \phi \b}$. With these expansions, we define a 
  new tunneling operator
\begin{equation}
\hat{\cal T}_{p,n}(v_p,\phi) = e^{-\phi^2/2} J_n(v_p/\wx) e^{i\phi
  \bbd} e^{i\phi \bb}\, \d.
\end{equation}

In order to illustrate the relevance of this term in the dynamics of
the system, we project the above expression onto the basis $\lbrace
|k,q\rangle \rbrace$, with $|k\rangle$ and $|q\rangle$ being the eigenstates of the 
 bosonic and the fermionic number operators, respectively. This means  $\hat{n}=\bbd\bb$ and 
 and  $\hat{n} |k\rangle = k |k\rangle$, and likewise, 
   $\hat{n}_e = \dd\d$,  and $\hat{n}_e |q\rangle = n_e | q\rangle$. Note that $n_e = 0 $ or $1$ for $|q\rangle=|-\rangle$(unoccupied) or
$|+\rangle$(occupied), respectively. Thus, the non-vanishing matrix
elements of the projected tunneling operator read
\begin{widetext}
\begin{eqnarray}\label{eq:tn}
\langle k- | \hat{\cal T}_{p,n}(v_p,\phi) |l+ \rangle &=&
J_{n}(v_p/\wx) \left( \sqrt{\frac{h_{k,l}!}{h_{l,k}!}}
\,e^{-\phi^2/2}\, (i\phi)^{|k-l|} \, L_{h_{k,l}}^{|k-l|}(\phi^2)
\right), \nonumber \\
\end{eqnarray}
\end{widetext}
with $L_n^\alpha (z)$ denoting the generalized Laguerre polynomials of
degree $n$, and
\[h_{k,l} =  k + \left(l-k \right)\Theta(k-l) .\] 
The magnitude of the tunneling matrix elements in Eq.~\eqref{eq:tn} are limited by the bounds of
the Bessel functions $\left| J_k(v_p/\wx) \right| \leq c/{v_p}^{1/3}$ with $c =
0,7853...$ \cite{Abramowitz1965}, and 

\[ \left| \sqrt{\frac{h_{k,l}!}{h_{l,k}!}} \,e^{-\phi^2/2}\, (i\phi)^{|k-l|} \,  L_{h_{k,l}}^{|k-l|}(\phi^2) \right| \leq 1 . 
\]

We emphasize that under these conditions, the weak coupling limit is valid 
also in the laboratory frame. Note that the structure of the {\it projected} tunneling term
Eq.~\eqref{eq:tn} is not in conflict with the assumption of the electronic lifetimes in 
the beam $\tau'$ to be longer than the typical time
scale $2\pi/ \omega_0$. Hence, the level broadening $\Gamma_n
\propto \sum_p|\hat{\cal T}_{n,p}|^2 $ is still small and the weak coupling
regime $\Gamma \ll \omega_0$ is realized for even moderate
values of $V_0$ and $\phi$.

\subsection{Rotating wave approximation}
\label{sec:effHam}

The quantum dynamics of the nanobeam for the choice of parameters 
$v_{l/r} \ll \omega_0$, $\phi \ll 1$, $\nu \ll
\omega_0 $, and small detuning $\delta\omega \ll \omega_0$, with 
\begin{equation}
  \delta \omega \equiv \omega_0 - \wx \, ,
  \end{equation}
 is most conveniently described in the co-rotating frame of reference, for which we apply a further unitary transformation
\begin{equation} \label{rotframe}
  R(t) = \exp[-i\, (\hn+ \dd \d ) \,\wx t] \, .
\end{equation}
In this rotating frame, the typical time scale of the system dynamics
is given by $\delta\omega^{-1}$, such that terms oscillating with
frequencies $\pm n\wx$ for $n\ge 1$ are averaged out and may be neglected in
the transformed Hamiltonian $H'(t) = R(t) H(t) R^\dagger(t) - i
R(t) \dot{R}^\dagger (t)$. For instance, applying the transformation of Eq.\ (\ref{rotframe}) to 
the quartic term in Eq.\ (\ref{hamm}), we obtain  
$R(t)(\bb+\bb^\dagger)^4 R^\dagger (t) = 6\hat{n}(\hat{n}+1)+O(\hat{b}^2 \exp[-2i\omega_{\rm ex} t])+
O(\hat{b}^{\dagger 2} \exp[2i\omega_{\rm ex}t])+3$, and in an appreciable amount of time, 
the terms $\propto \exp[\pm i n \omega_{\rm ex}t]$ for $n = 2$ and $4$ will quickly
average to zero, such that the relevant term in the quartic potential is $6\hat{n}(\hat{n} + 1) + 3$.

Within this rotating wave approximation (RWA), we find the Hamiltonian
\begin{equation}
\label{eq:efH}
      H' = H'_e + H'_m + H'_t + H'_{leads} ,
\end{equation}
where
\begin{eqnarray}
  H'_e &=& \delta \omega \, \dd \d \\[2mm] H'_m &=& \delta \omega \,
  \hn + \nu\, \hn (\hn + 1)/2 , \\[2mm]\label{eq:finaltunn} H'_t &=&
  \sum_{p,k} \hat{\cal T}'_p(v_p,\phi) \, \hat{c}_{k,p}^\dagger + {\rm
    h.c.}
\end{eqnarray}
Here, the tunneling operator
\begin{widetext}
\begin{equation}\label{eq:boson}
  \hat{\cal T}'_p (v_p,\phi) = T_p e^{-\phi^2/2}\sum_{nm} J_n(v_p/\wx)
  \frac{(i \,\phi)^{2m + n -1}}{m!(n+m-1)!} \, (\bb^{\dagger})^m \,
  \bb^{n+m-1} \d,
\end{equation}
\end{widetext}
has the matrix elements $\langle k - | \hat{\cal T}'_p(v_p,\phi) | l +
\rangle = \langle k -| \hat{\cal T}_{p,l-k+1} (v_p,\phi) | l +\rangle$
(cf. Eq.~\eqref{eq:tn}). In passing, we note that both unitary transformations given in Eqs.\ (\ref{eq:Ftrans}) and (\ref{rotframe}) commute with each other in the weak coupling regime in which the sequential tunneling approximation made below holds. Clearly, $R(t)$ as well as $\mathcal{U}(t)$  induce higher-order coupling terms within the transformed tunneling Hamiltonian which are beyond the sequential tunneling approximation used here. Moreover, for the tunneling term, the criteria of fast oscillating terms used in the rotating wave approximation is not sufficient to state that the contribution given in Eq.\ \eqref{eq:boson} is dominant over the neglected counter-propagating terms, and so the validity of the approximation needs to be verified. In the rotating frame, the tunneling term can be written as  
\begin{equation}\label{eq:fulltunn}
	\hat{\cal T}_p^{\, r}(v_p,\phi,t) = \sum_n \hat{t}_{n,p}(v_p,\phi) \, e^{in\wx t} \, ,   
\end{equation}	
 where $\langle k-|\hat{t}_{n,p}(v_p,\phi)| l+ \rangle =  \langle k -| \hat{\cal T}_{p,l-k+1+n} (v_p,\phi) | l +\rangle$. Thus for a small bias voltage, characterized by $v_p/\wx \ll 1$,  the ratio 
 between the matrix elements of the counter-propagating terms (for $n>0$ in Eq.\ \eqref{eq:fulltunn}) and the co-propagating ones ($n=0$) reads as
 \begin{equation}\label{eq:tunnratio}
 	\frac{\langle k-| \hat{t}_{n,p}(v_p,\phi)| l+ \rangle}{\langle k-|\hat{t}_{0,p}(v_p,\phi)| l+ \rangle}  \sim \frac{\Gamma(l-k+1)}{\Gamma(l-k+1+n)} \left( \frac{v_p}{2 \wx }  \right)^n \ll 1 \, .
 \end{equation}
Consequently, the contribution of the counter-propagating terms is negligible in the solution of the system dynamics, and the rotating-wave approximation, in which $\hat{\cal T}'_p(v_p,\phi) \equiv \hat{t}_{0,p}(v_p,\phi) $, is justified.  
 
\section{Quantum master equation}
\label{sec:masterequation}
The dynamics of the system described by the Hamiltonian
\eqref{eq:efH} is fully characterized by the statistical operator
$W(t)$, whose time evolution is governed by the von-Neumann equation
\begin{equation}
  \frac{d}{dt} W(t) = -i [H(t), W(t)]\, .
\end{equation}
After tracing out the degrees of freedom of the leads, we obtain 
the reduced system, represented by the density
operator $\rho \equiv {\rm Tr}_{\rm leads} [ W ]$.  In addition, in
the weak coupling regime considered throughout this work, $\Gamma \ll
\omega_0$, it is possible to express the evolution of the reduced
density operator in terms of a diagrammatic expansion in the tunneling
terms $H_{p}^{\pm}$ \cite{Weiss2015,Konig1996a}.  We use the 
standard Born-Markov approximation (for a recent discussion, see Ref.\ \onlinecite{Dubi2017})
and, furthermore, exploit a high-frequency approximation which is valid 
when the ac-voltage drive is much faster than the mechanical oscillations. Then, 
the master equation for the reduced
density operator reads
\begin{equation} \label{eq:master}
  \frac{d}{dt} \rho (t) = -i[H_0, \rho(t) ] + \Sigma \cdot \rho(t) \ ,
\end{equation}  
in which the first term on the right hand size represents the nanobeam
coherent dynamics, with
\begin{equation}\label{eq:H0}
H_0 = \delta \omega \, d^\dagger d + \delta \omega \, \hn
+ \nu \, \hn \, ( \hn + 1 ) / 2 \ ,
\end{equation}
and the second term represents dissipation and decoherence induced by
tunneling events between leads and the beam. This part is covered
by the self-energy $\Sigma =\Sigma_r+\Sigma_l$, therein $\Sigma_p$ is
the contribution from lead $p$.  In leading order in $\Gamma$, the
self energy $\Sigma_p$ is composed by eight terms corresponding to
different tunneling events $\Sigma_p = \sum_{n=1}^8\Sigma_p^{(n)}$
(see Appendix A for details).

\subsection{Electron current in the rotating frame}

The electronic current operator $\hat{I}_p$ from the lead $p$ to the nanobeam is given by 
the charge in the number of electrons in lead $p$ over time. We use the number operator of lead $p$ as $\hn_p = \sum_k \cc^\dagger_k \cc_k$ and find
\begin{eqnarray} \label{eq:current}
  \hat{I}'_p(t) &=& - e \dot{\hn}_p = -e [H'_t(t),\hat{n}_p] \nonumber
  \\ &=&\sum_k \left( \hat{\cal T}'_p(v_p,\phi) \cc^\dagger_{p, k}(t) \,
  - {\rm h.c.} \right)\ .
\end{eqnarray}
The net current passing through the nanobeam is 
\begin{equation} \label{eq:currentrwa}
 \hat{I}'_{\rm rwa} (t)\equiv
 \hat{I}'_l (t)- \hat{I}'_r (t)	\, .
\end{equation}
 We are left with calculating the expectation value
$I'_p(t)=\langle \hat{I}_p \rangle_t \equiv {\rm Tr} [
  \rho(t) \hat{I}_{p} ] $, which is determined from the self-energies and for which we find
\begin{equation}\label{eq:currenop}
  I'_p (t)= -i e \left\langle
  \Sigma_{p}^{(5)}(t)+\Sigma_{p}^{(8)}(t) -
  \Sigma_{p}^{(6)}(t)-\Sigma_p^{(7)}(t) \right\rangle .
\end{equation}
%

\subsection{Electron current in the laboratory frame}
In the previous section, the electron current has been expressed in terms of the self-energies 
$\Sigma_p^{(5),(8)}(t)$ and $\Sigma_p^{(6),(7)}(t)$ in the rotating frame. For a better
interpretation, we consider in this section the current in the laboratory frame.

Since the Hamiltonian is periodic in time, we can expand the diagrams in terms of 
 Fourier vectors $|f_m\rangle$, $m \in \mathbb{Z}$, such that 
 $\langle t|m\rangle = e^{-im \omega_{\rm ex} t}$. With this, the 
expectation value of the current becomes 
\begin{equation}
\label{eq:tcurrent}
I(t) = \sum_{m= -\infty}^{\infty} I_m \, e^{- i m	\omega_{\rm ex} t} 
\end{equation}
with the Fourier coefficients $I_m = I_{l,m} - I_{r,m}$, where 
\begin{align}
\label{eq:cfourier}
I_{p,m} &= -i e \left\langle \Sigma^{(5)}_{p,m}(\omega_{\rm ex}) +
\Sigma^{(8)}_{p,m}(\omega_{\rm ex})\right.\nonumber\\
& \left.- \Sigma^{(6)}_{p,m}(\omega_{\rm
  ex}) - \Sigma^{(7)}_{p,m} (\omega_{\rm ex}) \right\rangle \, .
\end{align}
Here, $\Sigma^{(i)}_{p,m}(\omega_{\rm ex})$ denotes the $m$-th Fourier
component of $\Sigma^{(i)}_p(t)$, i.e., $\Sigma^{(i)}_p(t) =
\sum_{m=-\infty}^\infty \Sigma^{(i)}_{p,m}(\omega_{\rm ex}) e^{-im \wx
  t}$.

The stationary value for $m=0$ and the higher harmonics ($m\neq 0$) of the current
are associated to different single-electron tunneling processes.  The
modulation in the bias voltage splits the energy levels of the leads
into sidebands separated by $\wx$.\cite{Tien1963} Thus, a lead state
$|k \rangle_{p}$, on lead $p$ and with energy $E_{k,p}$, is split into a set
of states $\lbrace |k_n \rangle_p \rbrace$ with energies $E_{nk,p} =
E_{k,p} + n \wx$, where $n$ is an integer and determines the order of the sideband.

In the undriven case,  the transfer of an
electron from the left to right lead occurs via  an energy 
level $E$ in the corresponding transport window characterized by 
$\mu_r < E <\mu_l$, where $\mu_p$ is the corresponding electrochemical potential of the lead $p$. 
On the
other hand, for the driven case, the condition for sequential
transport is not straightforward, since the occupied states on the
right lead can be above the Fermi level, i.e., there exists an $n$ such that
$E_{nk,r} > \mu_r$ although $E_{k,r} < \mu_r$ and a reduction of the
electronic current is generated.  Another interesting mechanism occurs
when an occupied sideband level on the left lead can reach the level
energy of the central system. This occurs when $E_{nk,l} = E$ for $E_{k,l} \neq
E$, and the electron can tunnel to the right lead to a sideband level
of energy $E_{kn',r} = E$ for $E_{k,r} > \mu_r$.  There, an electron
can transport $|n-n'|$ quanta of energy absorbed from the external modulation.
The current $I_p $ is given by the sum over all the possible tunneling
events from sidebands on the left lead to sidebands on the right lead.
If we denote by $P_{n n'}$ the probability of a tunneling event from
the sideband $|k_n\rangle_l$ to the sideband $|k_{n'}\rangle_r$, the
current can be rewritten in the form 
\begin{equation}
  I_p = e \sum_{nn'} P_{n n'} = \sum_{j=-\infty}^\infty \left( \sum_{n-n' = j}  e P_{n,n-j}\right).
\end{equation} 
This sum can be reordered according to the number of quanta of energy
exchange between the leads. Then, $\sum_{n-n'=m} eP_{nn-m}$ resembles the
aforementioned component $I_{p,m}$ and the stationary current
corresponds to $\sum_{n-n'=0} eP_{nn}$.

We are interested in the current for one-phonon processes characterized by 
\begin{equation}
I_{\rm rwa} (t) = I_{-1} \exp[i \wx t] + I_{1} \exp[-i \wx t] \, ,
\end{equation}
 whose oscillation amplitude in leading order of $v_p$ and $\phi$ corresponds to the current in the rotating frame, i.e.,
\begin{equation} \label{finres}
I_{\rm rwa } \equiv {\rm max}_t \, I_{\rm rwa} (t) = 2 |I_1| \, .
\end{equation}
\section{Quantum antiresonances}
\label{sec:quantumosci}
\begin{figure*}[ht]    
	\centering
        \includegraphics[width=0.8\textwidth]{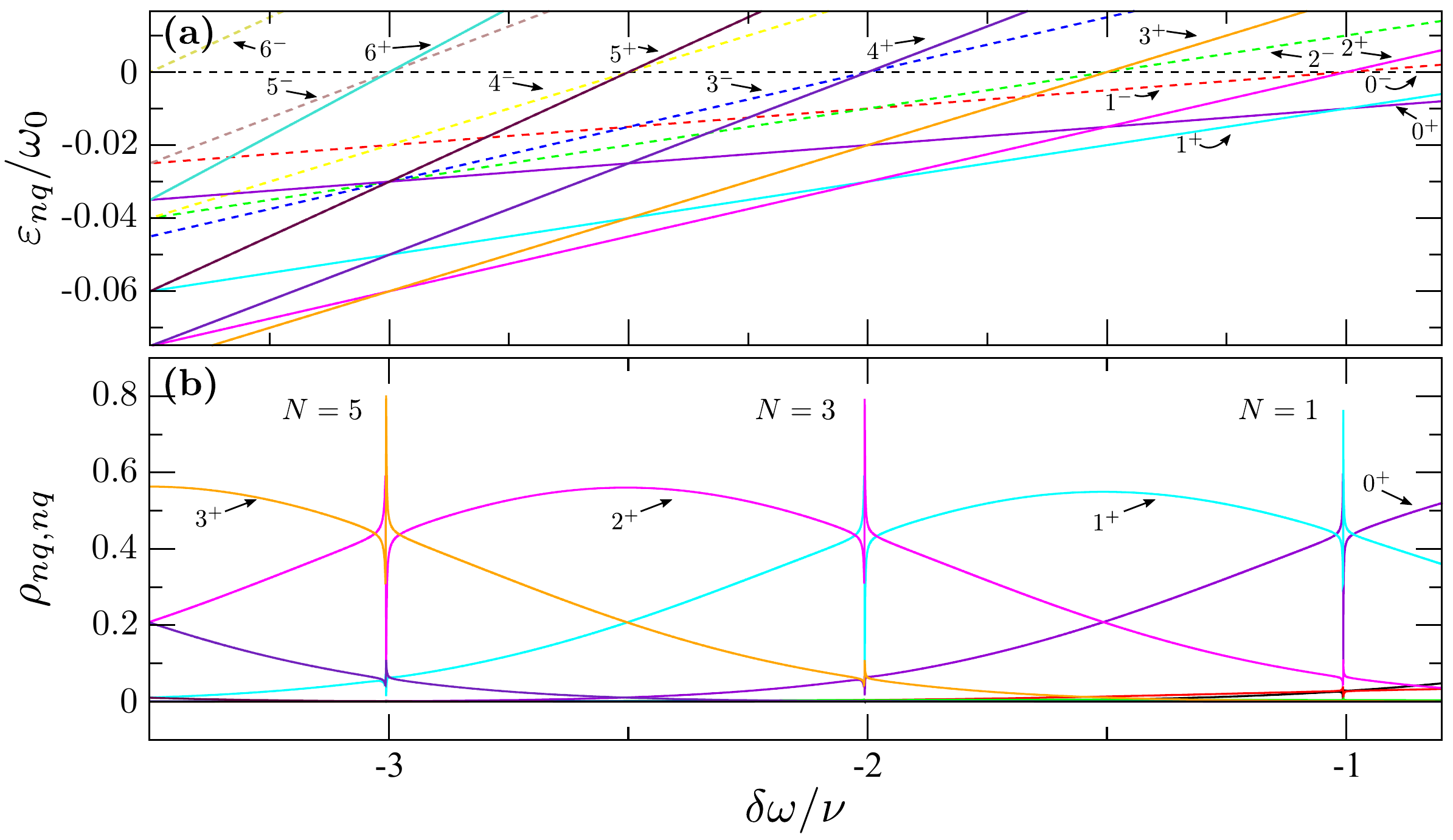} \caption{ \label{fig4}
          (a) Quasienergy spectrum for $\phi= 0 $ and (b) state
          populations (diagonal elements of the reduced density
          operator) for $\phi = 5 \times 10^{-2}$ as a function of the
          detuning frequency.  The labels $0^-, 1^-,
          2^-,...,6^-$ (dashed lines) mark the states for the
          electronically unoccupied nanobeam corresponding to the
          quasienergies $\varepsilon_{n-}$ in (a) and the populations
          $\rho_{n-n-}$ in (b) for $n = 0, 1, ... ,6$. Likewise, the labels
          $0^+,1^+,2^+,...,6^+$ (continuous lines) mark the
          electronically occupied nanobeam corresponding to the
          quasienergies $\varepsilon_{n+}$ in (a) and the populations
          $\rho_{n+;n+}$ in (b) for $n=0,1,2,...,6$.  The parameters
          used in these simulations are $ \Gamma = 10^{-3} \omega_0$,
          $\nu = 10^{-2} \omega_0$, $E = \omega_0$, $ V_l = - V_r =
          10^{-2}\omega_0$, $\mu_l - \mu_r = 10 \, \omega_0$, and $\kb
          T = 5 \times 10^{-3} \omega_0$. }
\end{figure*}

In general, switching off adiabatically the electromechanical coupling
$\phi \rightarrow 0$, mechanical and electronic subsystem decouple
and the energy levels of the nanaobeam
increase by multiples of the mechanical nonlinear strength $\nu$ according to Eq.~(\ref{eq:H0}) as 
\begin{equation}
\varepsilon_{n+1,q} - \varepsilon_{n,q'} = \delta\omega (1+n_q-n_{q'})
+ \nu (n +1)\, .
\end{equation} 
We label the eigenstates by $|n,q \rangle$, where $n
= 0,1,2, \dots$ denotes the quantum number of the vibrational state and $q =+,-$ refers to the
quantum number of the electronic state, respectively. Since we use a spinless model, we have two possible electronic eigenstates of the dot, either the occupied or the unoccupied state for the occupation number operator $n_q=d^\dag d$. Suppose that $N$ quanta of energy have been exchanged between the two subsystems, then several non-equidistant resonances will appear in the spectrum. They are quantified by the quasienergy $\varepsilon_{N-n,q} = \varepsilon_{n,q'}$, i.e., the detuning should be chosen as
\begin{equation}\label{eq:cond}
\delta\omega = - \frac{\nu}{2}\,\, \frac{N(N+1)}{N + n_q - n_{q'}} \, .
\end{equation}
%
Note that for the nontrivial case $N>0$, the detuning is always negative $\delta \omega<0$.  In Fig.~\ref{fig4}(a), the quasi-energy spectrum, for
$\phi = 0$, as a function of the ratio $\delta \omega /
\nu$ is shown. Exact crossings occur whenever the condition of Eq.~\eqref{eq:cond} is met, indicating a degeneracy between two quasienergies. Note that for the linear 
Holstein model \cite{Weiss2015}, when $\nu=0$, all degeneracies are absent. For $\phi \neq 0$, the degeneracy is lifted and the states
$|N-n, q \rangle$ and $|n, q' \rangle$ are mixed by the interaction terms, see Eq.~\eqref{eq:finaltunn}. They generate the anticrossings of the quasienergy levels in the spectum. 
Around a given (anti-)resonance, the states $|N-n,q\rangle$ and $|n,q'\rangle$ are mixed strongest (see
Fig.~\ref{fig4}(b)).  The mixing results in the corresponding dressed
states $|\varphi_n\rangle$ and $|\varphi_{N-n}\rangle$, which are 
superpositions of the two localized states $|N-n,q\rangle$ and
$|n,q'\rangle$.  As an analogy, one might think of a static double-well potential, where for a finite overlap between the two degenerate states (referred to as tunneling), the left and right energy eigenstates are mixed and  the spectrum forms an anticrossing when the bias between the two wells is changed.  
Here, we would identify the left and right localized states with the pairwise resonant  states $|N-n,q\rangle$ and
$|n,q'\rangle$. Naturally, the role of the 'tunneling' is played by the electromechanical coupling $\phi$, which  induces a coupling between the two states and thus generates transitions.

In the laboratory frame, the electrons couple to the mechanical motion
via the operator $\bbd - \bb$, see Eq.~(\ref{eq:elemech}). This
means that the mechanical degree of freedom receives or releases energy, once the electronic
state is occupied.  Therefore, the most populated states are formed by 
the pair $|N-n,+\rangle$ and $|n,+\rangle$. This behavior is in
analogy to the Duffing oscillator \cite{peanoPRB,peanoCP,peanoNJP}, where at resonance the population
is concentrated on those states, i.e., $\rho_{N-n,+} = \rho_{N,+} =
1/2$.

On the other hand, the expansion used here, in leading order of 
$\Gamma$, considers transitions between nearest
neighbor states. The transition dynamics between states of the 
vibrating nanojunction affects the sequential
tunneling current when $N$ is an odd integer. In the picture of a bistable quasienergy surface 
\cite{peanoNJP}, this amounts to a single phonon
inter-well transition. The nearest-neighbor condition on
$|\varphi_{(N-n)+}\rangle$ and $|\varphi_{n+}\rangle$ requires that
$(N-n) - n = 1$, such that $N$ is an odd number.  Thus, in the case of the 
$N$-th resonance with $N$ being odd, the relevant states are $|N^*+1,+\rangle$ and
$|N^*,+\rangle$ with $N^* = (N-1)/2$.

Below, we aim at obtaining an approximate and simple expression for the line shape of the 
antiresonance in the current. For this, we need the approximate solution of the quantum master equation in 
the vicinity of an avoided crossing of a pair of quasienergy states. 
In leading order of the voltage, i.e., of the ratio $v_p/\wx$, and of $\phi$, we keep the terms for $(n,m)=(0,1)$
and $(n,m) = (1,0)$ in the tunneling operator in Eq.~(\ref{eq:boson}), which yields to  
 a simplified expression in the form  
\begin{equation}\label{eq:tun3}
  \hat{\cal T}'_p(v_p , \phi) \approx \frac{v_p}{2 \wx} \mathbb{I}_{\rm
    mech} \d + i \, \phi \, \bbd \d \, .
\end{equation}
We used $J_1(v_p/\wx) \approx v_p/(2\wx)$ and $J_0(v_p/\wx)
\approx 1$. Above, $\mathbb{I}_{\rm mech} = \sum_n |n\rangle \langle n|$ is
the identity operator in the Hilbert subspace of the mechanical degrees of freedom with the basis
$\lbrace |n\rangle \rbrace$. The above expression Eq.~\eqref{eq:tun3}
yields the self-energies
\begin{eqnarray} \label{eq:self1}
  {}^{n +}_{n+} \left[ \Sigma_p^{(5)}+\Sigma_p^{(8)} \right]_{m -}^{m -}
  &\approx& (n+1)\, \phi^2\, \Gamma_p \, f^-_p[- \nu(n+1)]
  \,\delta_{m,n+1} \nonumber \\ &&\\ \label{self2} {}^{m-}_{m-} \left[
    \Sigma_p^{(6)}+\Sigma_p^{(7)} \right]_{n+}^{n+} &\approx& (n+1)\,
  \phi^2\, \Gamma_p f^+_p[-\nu(n+1)] \,\delta_{m,n+1}\nonumber \\ 
\end{eqnarray} 

where $f_{p}^\pm (\epsilon) = \{\exp[\pm(\epsilon -
  \mu_p)]+1\}^{-1}$ is the probability distribution of an   
  occupied ($+$) or an unoccupied ($-$)
electronic state in the lead $p$.  The self-energies
Eqs.~\eqref{eq:self1} and \eqref{self2} represent transition rates between 
different mechanical states which are relevant for the current
calculation. It follows that
the coupling with the leads only induces transitions between nearby
mechanical states within this single-phonon approximation.

The external bias voltage modulation induces a transition 
from $|N^*, +\rangle $ to $|N^*+1, +\rangle$,  
while electron tunneling generates transitions 
between nearby mechanical states.  
As a consequence, the ratio $\rho_{n+;n+}/\rho_{n+1,+;n+1,+}$ is given 
by the ratio of the corresponding transition rates as 
\begin{equation}
  \frac{\rho_{n+;n+}}{\rho_{n+1,+;n+1,+}} = \frac{\quad \quad \quad
    {}^{n +}_{n+} \left[ \Sigma^{(5)}_p + \Sigma^{(8)}_p
      \right]_{(n+1) -}^{(n+1) -}}{{}^{(n+1) +}_{(n+1)+} \left[
      \Sigma^{(5)}_p + \Sigma^{(8)}_p \right]_{n -}^{n -} } \approx
  \frac{n+1}{n+2} \, .
\end{equation}
Taking into account that $\rho_{N^*+;N^*+} = \rho_{N^*+1,+;N^*+1,+}$,
the states $|\varphi_{N^*+}\rangle $ and $|\varphi_{N^*+1,+}\rangle $
are the states with the largest occupation probability.

To summarize, the electrons on the leads exchange energy with the
external modulation, thereby getting dressed. Then, a dressed electron tunnels to the central
system sending the mechanical motion out of equilibrium due to the
electromechanical coupling. Depending on the external frequency, the
mechanical motion can exchange energy with the electrons affecting the
amplitude of the electronic current. This provides feedback to the current. 
This process is similar to controling the thermal occupation of the vibrational mode of magnetic
\cite{Brueggemann2014,Brueggemann2016} and non-magnetic\cite{Weiss2015} molecular junctions 
by an external spin current. There, the magnetic moment and the vibrational mode interact via a magnetomechanical coupling, 
yielding to an exchange of energy in the way that the vibrational energy can be transferred to the  magnetic degree of freedom, which overall implies vibrational cooling of the nanojunction. 

In panel (b) of Fig.~\ref{fig4},
the steady-state populations of the system are depicted as a function of
the external frequency. The most populated states correspond to
$\rho_{N^*+1,+;N^*+1,+}$ and $\rho_{N^*+;N^*+}$. Out of resonance, 
all the states are equally populated and $\rho_{nq;nq} = 1/(2\bar{N})$,
with $\bar{N}$ being the number of states covered within the bias window
$\mu_l - \mu_r$. For the $N$th resonance when $N$ is even, the most populated
state is $\rho_{N/2,+;N/2,+}$ and $\rho_{N/2+1,+;N/2+1,+} =
\rho_{N/2-1,+;N/2-1,+}$ due to the single excitation process induced by
the current. Those states are dominant. Therefore, $\rho_{N/2,+;N/2,+} +
\rho_{N/2+1,+;N/2+1,+} +\rho_{N/2-1,+;N/2-1,+} \approx 1$ as it
is shown in the panel (b) of Fig.~\ref{fig4}.

\subsection{Signatures in the electronic current}

From Eqs.~\eqref{eq:self1} and \eqref{self2}, we obtain directly an 
analytic approximation for the electronic current in the rotating frame (cf. Eq. \eqref{eq:currenop}) 
\begin{widetext}
\begin{equation} \label{eq:current2}
  I'_p(v_p , \phi) \approx \phi^2\, \Gamma_p \, \sum_n (n+1) \left(
  f^-_p[-\nu(n+1)] \rho_{n+;n+} - f^+_p[-\nu(n+1)] \rho_{n-;n-} \right) .
\end{equation} 
\end{widetext}
The first (second) term on the right hand side in
Eq.~\eqref{eq:current2} corresponds to the current of out-coming
(incoming) electrons from (into) the central system, respectively.
Around the $N$th resonance when $N$ is odd, the populations $\rho_{N^* +; N^*+}$ and
$\rho_{N^*+1, +;N^*+1, +}$ are dominant 
and the current simplifies to

\begin{eqnarray}\label{eq:currentapprox}
  I'_p(v_p , \phi) &\approx& \phi^2\, \Gamma_p \, (N^*+1)
  f^-_p[-\nu(N^*+1)] \nonumber \\ && \quad \quad \times \left(
  \rho_{N^* +; N^* +} + \rho_{N^*+1, +;N^*+1,+} \right) . \nonumber
  \\ &&
\end{eqnarray} 
With this at hand, we can calculate the current amplitude $I_{\rm rwa}  = 2 |I_1|$ of Eq.\ (\ref{finres}) in the laboratory frame. 

\begin{figure*}[ht]    
	\centering
        \includegraphics*[width=0.8\textwidth]{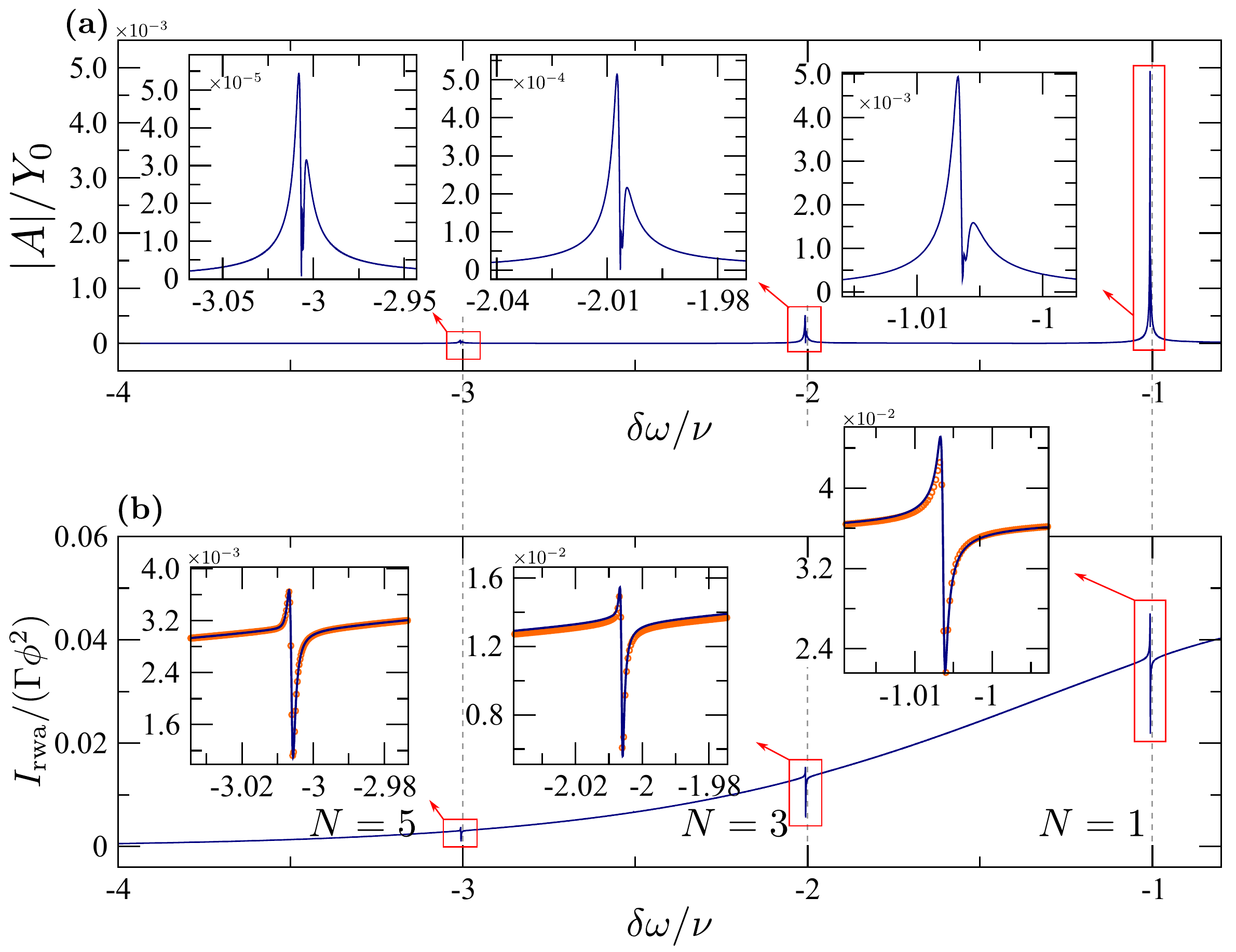}
	\caption{\label{fig2} (a) Amplitude of the mechanical vibration, 
          and, (b) the amplitude of the first harmonic of the ac current $I_{\rm rwa}=2 |I_{1}|$, Eq.\ (\ref{finres}),  as a function of the detuning
          frequency $\delta \omega$. The parameters used in these
          simulations are $\phi = 5 \times 10^{-2}$, $ \Gamma =
          10^{-3} \omega_0$, $\nu = 10^{-2} \omega_0$, $E = \omega_0$,
          $V_l = - V_r = 10^{-2} \omega_0$, $\mu_l - \mu_r = 10 \,
          \omega_0$, $k_BT = 5 \times 10^{-3} \omega_0$. 
         }
\end{figure*}

Following a similar procedure used for the calculation of the current \eqref{eq:currentapprox}, we  write the tunneling operator $\hat{t}_{n,p}(v_p,\phi)$ in leading order of $v_p$ and $\phi$ and calculate the relevant self-energies $\Sigma_p^{5,8}$ and $\Sigma_p^{6,7}$. Then, the current amplitude for the $n$-phonon process, for $n>1$, is given by $I_n = J_{n-1}^2(v_p/\wx) I_{\rm rwa}$. Consequently, $I_n/I_{\rm rwa} < (v_p/\wx)^{2n-2} \ll 1$, which shows that the contribution from the co-rotating terms are also dominant for the current. The amplitude $I_{\rm rwa}$ of the current 
flowing through the central system is shown in panel (b) of Fig.~\ref{fig2} as a function of the detuning of the external driving 
frequency (blue continuous line). The blue solid lines indicate the 
calculated full mean value Eq.~\eqref{eq:currentrwa} without further approximation.  We find pronounced antiresonances at particular values of the detuning. The current describes an asymmetric line-shape
resonance determined by the resonance condition established in 
Eq.~\eqref{eq:cond} for $N$ odd. Inside panel (b), a zoom of the current
behavior around the antiresonance is shown. 
In addition, we also show the current calculated using the approximation 
Eq.~\eqref{eq:currentapprox} (orange dots). Both results agree well which underlines 
that the argumentation yielding us to the approximation is correct. The antiresonances are similar to those 
obtained for the dissipative quantum Duffing oscillator \cite{peanoPRB,peanoCP,peanoNJP} and those of the 
driven dissipative Jaynes-Cummings model \cite{peanoJCM1,peanoJCM2} and they have a Fano-type form due to the fact that a discrete quantum level interacts with a continuum of electronic energy levels. 

In this regime, in which $v_{p}/\wx \ll \phi$, the time-averaged
input-output power $P(\wx) = \wx \langle d \hat{n} /dt \rangle $ is
proportional to the electron current \eqref{eq:currentapprox}, i.e., 
\begin{eqnarray}\label{eq:power}
  P (\wx) &=& - i \wx \langle \left[ H'_t, \hat{n} \right] \rangle
  \nonumber \\ &\approx& -i \wx \sum_p p\langle
  \Sigma_p^{(5)}+\Sigma_p^{(8)} - \Sigma_p^{(6)}+\Sigma_p^{(7)}\rangle.
  \nonumber \\ &&
\end{eqnarray}

This means in turn that measuring the electrical current gives insight
into the population of the mechanical states and the flux of excitations 
put into the motion of the clamped beam.

\subsection{Antiresonant mechanical nonlinear response}
According to Eq.~\eqref{eq:power}, the electronic current drives the
mechanical degree of freedom out of equilibrium.  An interesting consequence for the
mechanical motion is the nonlinear response of the nonlinear nanobeam to the
external ac driving of the bias voltage. We are thus interested in the
nonlinear response of the mechanical motion characterized by the mean value $A$ 
of the position operator in the steady state, defined by 

\begin{equation}
  A = \frac{Y_0}{\sqrt{2}}\; {\rm Tr} [ (\bb + \bb^\dagger ) \rho ] \, .
\end{equation}   
Here, $Y_0$ is the amplitude of the zero point fluctuations in the
nanobeam's fundamental  bending mode. Note that this mean value
corresponds to the oscillation amplitude of the expectation value of the
position operator in the laboratory reference frame. Therefore, we
denote $A$ as the amplitude of the nonlinear response.

Around the $N$th resonance ($N$ odd), the quasi-energy difference between
$|\varphi_{N^*,+} \rangle$ and $|\varphi_{N^*+1,+} \rangle$ becomes 
smaller than $\Gamma$. We can consider $|\varphi_{N^*,+} \rangle
\approx (|N^*,+\rangle + |N^*+1,+\rangle)/\sqrt{2}$ and $|
\varphi_{N^*+1,+} \rangle \approx (|N^*,+\rangle -
|N^*+1,+\rangle)/\sqrt{2}$. With this, we find 
\begin{equation}
  A \approx \frac{Y_0}{\sqrt{2}} \sqrt{N^* + 1} \left(
  \rho_{N^*,+;N^*,+} - \rho_{N^*+1,+;N^*+1,+} \right)\, .
\end{equation}
The contribution from off-diagonal elements add up to zero due to
$\rho_{N^*,+;N^*+1,+}=\rho^*_{N^*,+;N^*+1,+}$.

At resonance, each state of the corresponding pair has the same occupation
probability, $\rho_{N^*,+;N^*,+} = \rho_{N^*+1,+;N^*+1,+}$, and the
nonlinear response amplitude vanishes $A=0$.  Away from resonance the
pair-wise states are localized, $|\varphi_{N^*,+} \rangle =
|N^*,+\rangle$ and $|\varphi_{N^*+1,+} \rangle = |N^*+1,+\rangle$,
and their quasi-energy difference is larger than the tunneling constant 
$\Gamma$. Therefore, the off-diagonal elements of the density matrix
are negligible, yielding $A = 0$. In Fig.~\ref{fig2} (a), the
amplitude of the nonlinear response is depicted as a function of the
external frequency. Again, the amplitude exhibits quantum antiresonances
 around the $N$th resonance ($N$ odd). Since this antiresonance appears 
 in the current spectrum, it should in principle be directly measurable. 

\section{Conclusions}
\label{sec:conclusions}
The interplay of a dissipative nonlinear quantum mechanical resonator with an external periodic driving 
is known to generate nontrivial response properties of the resonator  in the 
form pronounced and rather sharp quantum antiresonances. The detection of those 
is non-trivial. In this work, we proposed to use a molecular nanojunction (or, a nanobeam) in its 
regime of nonlinear mechanical oscillations and clamped to conducting leads. This junction carries electronic current when an ac driving voltage is applied. An applied static magnetic field controls 
the electromechanical coupling of the flowing electron current and the mechanical 
oscillation. A static
longitudinal compression force close to the Euler buckling instability may 
be used to tune the nonlinearity. Then, the mechanical oscillation amplitude can
be described by an effective single particle quantum harmonic oscillator
Hamiltonian with a weak Kerr nonlinearity.  For the electronic part, we consider
weak tunneling contact between the junction and the lead.  
The first longitudinal energy eigenstate is
associated with the motion of the electrons along the nanobeam, such that 
a quantum dot is formed. 

In the regime of weak electromechanical coupling, and considering a
finite lifetime of the electrons in the nanojunction being longer than the typical
time scale of intrinsic junction dynamics, the non-equilibrium dynamics is captured by
a Born-Markov master equation. It has been formulated in a frame
rotating with the ac driving frequency, in which the fast oscillating
terms were average out. The effective model 
Hamiltonian of the molecule shows non-equidistant quasi-energy levels which 
define several resonant conditions which corresponds to multiquantum 
transitions in the nanobeam mechanical motion. In particular, the mechanical response 
reveals striking quantum antiresonances between pairs of quasienergy levels which 
for an anticrossing when, for instance, the driving frequency is varied. 
For modulation frequencies around the defined resonance conditions, the dynamics can
be simplified by restricting to a two quasi-energy levels only. The solution may 
be used to determine the flowing electron current which is the observable being 
directly accessible in an experiment. The approximate picture is confirmed by solving
the full master equation numerically and by calculating the net current passing
through the nanobeam. Although we have presented results for a specific set of parameters in this work, especially the quantum master equation allows one to explore further regions of the parameter space. The observed effects will also survive in the regime of strong nonequilibrium quantum transport, where higher order phonon processes become important. 
We find that the feature of the quantum antiresonances in the 
mechanical response of the junction translates into antisymmetric line shape resonances 
in the electric charge current located at frequencies where the multiple
transition in the mechanical motion takes place.

For very weak driving amplitudes of the ac voltage, we find a simple expression for the
current which shows its direct dependence on the occupation
probability of the mechanical antiresonant states. Along with the
electronic current, a flux of energy into or out of the nanojunction can be determined, 
which drives the mechanical degree of freedom 
out of equilibrium. We find a similar structure
of the nonlinear response of the nanojunction with that
calculated for the quantum Duffing oscillator \cite{peanoPRB,peanoCP,peanoNJP}. Moreover, 
the response is also similar to the driven dissipative Jaynes-Cummings model 
\cite{peanoJCM1,peanoJCM2}. Yet, the important difference in the present quantum transport set-up is that 
the quantum antiresonances are directly measurable in the current which renders the effect 
interesting for experimental observation.  

\acknowledgments
V.L. was supported by the project 935-621115-N24 Universidad Santiago
de Cali, Colombia.

\appendix

\section{ Quantum master equation approach}
To obtain the dynamics of the central nanojunction only, it is convenient to trace out
the electrodes' degrees of freedom in the full density operator $W(t)$ of leads plus junction. In doing so in the interaction picture, the reduced density
operator reads 
\begin{eqnarray}\label{eq:rhointime}
{\rho}^I (t) &=& \Tr_{l,\, r} \{ {W}^I(t)\} \nonumber \\ &=& {\rho}(t_0)
-i \int_{t_0}^t dt_1 \Tr_{l,\,r} [ {H}^I_T(t_1), {W}^I(t_1) ] \ ,
\nonumber \\ &&
\end{eqnarray}
where $\Tr_{l,\, r}$ denotes the trace over the degrees of freedom of
the right and left lead.  Differentiating with respect to $t$, we
obtain the quantum master equation for the reduced density operator,
\begin{equation}\label{eq:diffrho}
\frac{d}{dt}  {\rho}^I (t) = -  \int_{t_0}^t
dt_1\ \Tr_{l,\,r} \left\{ [ {H}^I_T(t),[ {H}^I_T(t_1), {W}^I(t_1)]] \right\} \, ,
\end{equation}
where, for simplicity, we have eliminated the term $-i 
\Tr_{l,\,r} \bigl\{ [ {H}^I_T(t), {W}^I(t_0)] \bigr\}$ with the
assumption $\Tr_{l,\,r}\bigl\{ {H}^I_T(t) {\rho}_l\otimes {\rho}_r
  \bigr\} =0$.  This is equivalent to consider $ {\rho}_l\otimes
{\rho}_r$ as diagonal in energy basis, in other words, it is equivalent
to the assumption that the leads are at their respective thermal equilibrium.
${W}^I$ factorizes at $t = t_0$, and at later times correlations
between leads and the central system arise due to the tunneling
term $ {H}_T$. However, for a very weak coupling, at all times $
{W}(t)$ should only show deviations of order ${H}_T$ from an
uncorrelated state.  Thus, in this regime of sequential electron tunneling, 
 we can formulate a quantum master equation for
$\rho$ in the form 
\begin{equation}\label{eq:tme}
\frac{d}{dt}  {\rho} (t) =  -i [ {H}_0,  {\rho}(t)] +
\int_{t_0}^t dt_1 \, \Sigma(t,t_1) \cdot  {\rho}(t_1) \ .
\end{equation}
in the Sch\"odinger picture.  The first term on
the right hand side in Eq.~\eqref{eq:tme} governs the coherent dynamics,
whereas the second term encloses all the effects of the fermionic bath
covered by the kernel $\Sigma(t,t_1)$. It includes self-energies, which are induced by
the leads, in arbitrary orders in the tunneling.  In a diagrammatic
expansion of Eq.\ (\ref{eq:tme}) \cite{Konig1996a}, the selfenergy $\Sigma(t,t_1)$ encloses
only irreducible terms.

To calculate these irreducible diagrams, it is convenient to split the
tunneling term $ {H}_t$ into two parts, according to Eq.\ \ref{eq:tunneling}. 
In order to simplify the notation, we omit here the superscript for the
interaction picture for the creation and annihilation operators. It is
implicitly assumed unless stated otherwise. Then, 
the lowest order of the expansion of $\Sigma$ can be written
as
\begin{widetext}
\begin{eqnarray}
\label{eq:Sigma1}
&& \int_{t_0}^t dt_1 \Sigma(t,t_1) {\rho}(t) = \Sigma(t)
{\rho}(t) \nonumber \\ &=& -\T_K \int_{K} ds \left(
\underbrace{\left\langle {H}_t^+(t) {H}_t^-(s)\, {\rho}(t)
	\right\rangle_{lr}}_{\Sigma^{(1)}(t,s) \cdot \rho(t)}
+ \underbrace{\left\langle {H}_t^-(t) {H}_t^+(s)
  {\rho}(t)\right\rangle_{lr}}_{\Sigma^{(2)}(t,s) \cdot \rho(t)}
\right.  \nonumber \\ && \quad \quad \quad \quad \ \, + \;
\underbrace{\left\langle {\rho}(t) {H}_t^+(s) {H}_t^-(t)
	\right\rangle_{lr}}_{\Sigma^{(3)}(t) \cdot \rho(t)}
+ \underbrace{\left\langle {\rho}(t){H}_t^-(s) {H}_t^+(t)
	\right\rangle_{lr}}_{\Sigma^{(4)}(t,s) \cdot \rho(t)} \nonumber
\\[0.2cm] && \quad \quad \quad \quad \ \, - \;
\underbrace{\left\langle {H}_t^+(t) {\rho}(t) {H}_t^-(s)
	\right\rangle_{lr}}_{\Sigma^{(5)}(t,s) \cdot \rho(t)} -
\underbrace{\left\langle {H}_t^-(t) {\rho}(t) {H}_t^+(s)
	\right\rangle_{lr}}_{\Sigma^{(6)}(t,s) \cdot \rho(t)} \nonumber
\\[0.2cm] && \quad \quad \quad \quad \ \, \left.- \,
\underbrace{\left\langle {H}_t^+(s) {\rho}(t) {H}_t^-(t)
	\right\rangle_{lr}}_{\Sigma^{(7)}(t,s) \cdot \rho(t)} -
\underbrace{\left\langle {H}_t^-(t_1) {\rho}^I(t) {H}_t^+(t)
	\right\rangle_{lr}}_{\Sigma^{(8)}(t,s) \cdot \rho(t)} \right) \ ,
\nonumber \\[0.2cm] &&
\end{eqnarray}
\end{widetext}
where $\langle \cdots \rangle_{lr} = \Tr_{l,\,r}\{(\cdots) {\rho}_l
\otimes {\rho}_r\}$. $K$ denotes the closed Keldysh contour which
runs from $t_0$ to $t$ on the real axis and then back again from $t$
to $t_0$. Moreover, $\T_K$ denotes the corresponding time ordering operator on the
Keldysh contour.

The formal solution of the quantum master equation Eq.~\eqref{eq:master} can be
cast into the form 
\begin{eqnarray}\label{Tunnmaster}
{\rho}(t) &=& e^{{\cal L} \, t} {\rho}(0) = \sum_k 
\Tr\bigl\{ {\rho}_k^\dagger  {\rho}(0)\bigr\}\,e^{\Gamma_k t}\,
{\rho}^k \ , 
\end{eqnarray}
where $ {\cal L}  {\rho}(t) \equiv -i[ H_0 , \rho(t) ] +
\Sigma \rho(t)$, $ {\rho}^k$ ($ {\rho}_k$) are the right (left)
eigenoperators of $\L$ with eigenvalue $\Gamma_k$.  The steady state
is determined by the right eigenoperator $ {\rho}^{k'} \equiv
{\rho}^\infty$ with the eigenvalue $\Gamma_{k'} =0$. Therefore, the
solution of the master equation is linked to the solution of a
eigenvalue problem for a singular matrix.

\section{Charge current in the laboratory frame}
We may use the expansion of the quantum master equation
 to derive an expression for the electric charge current.  
 By definition, the current is
given by the time derivative of the electron number $ \hat{n}_{p} =
\sum_k {c}_{k,p}^\dagger {c}_{k,p}$ on lead $p$, i.e., by 
\begin{eqnarray}\label{current}
I_{p}(t) &=& - e \frac{d}{dt} \left\langle \hn_{p} (t)
\right\rangle = - i e \left\langle \bigl[ {H}(t),
\hn_{p}(t)\bigr] \right\rangle \nonumber \\ &=& - i e \left(
\bigl\langle {H}^+_{t,p}(t) \rangle - \langle {H}^-_{t,p} (t)
\bigr\rangle \right) \ .
\end{eqnarray}
In leading order of $\Gamma$, the current is determined by the
components of the self energy, $\Sigma_{p}^{(5...8)}$,

\begin{widetext}
\begin{eqnarray}\label{current3}
I_{p}(t) &=&  \sum_m I_{p,m}\, e^{i\, m \, \omega_{\rm
		ex}\, t} \, \with   \\ I_{p,m} &=& -i e \left\langle 
\Sigma^{(5)}_{p,m}(\omega_{\rm ex}) +
\Sigma^{(8)}_{p,m}(\omega_{\rm ex}) - 
\Sigma^{(6)}_{p,m}(\omega_{\rm ex}) -
\Sigma^{(7)}_{p,m} (\omega_{\rm ex})
\right\rangle \nonumber \, ,
\end{eqnarray}
\end{widetext}
where, in order to calculate the current, we fix \cite{Weiss2015} one tunneling vertex in each of the diagrams at the measurement time $t$. Here, $\Sigma^{(i)}_{p,m}$ is the $m$th Fourier component of
$\Sigma^{(i)}_{p}$.  We symmetrize with respect to the leads, such
that we compute the current $I(t) = I_l(t) - I_r(t)$.
In the rotating frame of reference, after neglecting the fast
oscillating terms, the calculated current correspond to the amplitude
of the first mode in Eq.~\eqref{current3}.

\end{document}